\documentclass[12pt,epsfig]{article}
\usepackage{amssymb,epsfig}

\newcommand{\ve} {\varepsilon}
\newcommand{\half} {\frac{1}{2}}
\newcommand{\quart} {\frac{1}{4}}
\newcommand{\abs} [1] {\vert#1\vert}
\newcommand{\im} {\mathrm{Im}}

\newcommand{\Ahat} {\widehat{A}}
\newcommand{\ps} {\left(\frac{p}{\alpha}\right)}
\newcommand{\qs} {\left(\frac{q}{\alpha}\right)}

\begin{document}

\begin{titlepage}
\hfill
\vbox{
    \halign{#\hfil         \cr
           } 
      }  
\vspace*{20mm}

\begin{center}
{\Large {\bf On the Entropy of Four-Dimensional Near-Extremal $N=2$
Black Holes with $R^2$-Terms}\\} \vspace*{15mm}

{\sc Eyal Gruss} \footnote{e-mail: {\tt eyalgruss@gmail.com}} and
{\sc Yaron Oz} \footnote{e-mail: {\tt yaronoz@post.tau.ac.il}}

\vspace*{1cm}
{\it School of Physics and Astronomy,\\Raymond and Beverly Sackler Faculty of Exact Sciences,\\
Tel-Aviv University, Tel-Aviv 69978, Israel.\\}

\end{center}

\vspace*{8mm}

\begin{abstract}
We consider the entropy of four-dimensional near-extremal $N=2$
black holes. Without $R^2$-terms, the Bekenstein-Hawking entropy formula has the
structure of the extremal black holes entropy with a shift of the
charges depending on the non-extremality parameter and the moduli at
infinity. We derive a generalized Wald entropy formula for non-extremal $N=2$ black holes with $R^2$-terms.
We construct a class of small near-extremal horizon solutions with $R^2$-terms.
In this case the entropy is the same as in the extremal case to order $O(\mu^2)$, and does not exhibit the charge shift behavior.

\end{abstract}
\vskip 0.8cm

August 2007

\end{titlepage}

\setcounter{footnote}{0}

\section{Introduction}

Explaining the microscopic origin of black holes entropy is one of
the important tasks of any theory of quantum gravity. Much
progress towards achieving this goal in the framework of string
theory has been obtained for supersymmetric charged black holes in
various dimensions. One such class of extremal black holes
characterized by electric and magnetic charges $(q_I,p^I)$ exists
in compactification of type II string theory on Calabi-Yau
$3$-folds ($CY_3$) (for a review see
\cite{mohauptreview,recentreviews}). These black holes are
obtained by wrapping $D$-branes around cycles in $CY_3$. Their near
horizon geometry is $AdS_2\times S^2\times CY_3$, where the moduli
of the Calabi-Yau $3$-fold are fixed on the horizon by the
attractor equation in terms of the charges. Recently much work has
been done on extremal non-supersymmetric black holes
\cite{nonsusy}.

The four-dimensional low energy effective action of type II strings
compactified on $CY_3$ is given by $N=2$ Poincar\'e supergravity
coupled to $N=2$ Abelian vector multiplets. The macroscopic extremal
black holes are asymptotically flat charged supersymmetric solutions
of the field equations. At leading order in the curvature, the
entropy of the black holes is given by the Bekenstein-Hawking area
law $S = \frac{A}{4}$, where $A$ is the area of the event horizon
and is determined in terms of the charges by the attractor
mechanism. With subleading $R^2$-terms included, the entropy of
these macroscopic black holes has been computed using the
generalized entropy formula of Wald \cite{wald}.

With one electric charge $q_0$ and $p^A$ magnetic charges
\footnote{We consider type II compactification with $A=1\ldots b_2$
and $b_2$ is the second Betti number of $CY_3$.} one gets the
entropy \cite{r2entropy}
\begin{equation}
\label{ent} S= 2\pi\sqrt{q_0\left(D_{ABC}p^Ap^Bp^C + 256
D_Ap^A\right)} \ , \label{entfor}
\end{equation}
where $D_{ABC}$ and $D_A$ are respectively proportional to the
triple intersection numbers and the second Chern class numbers of
the $CY_3$. The first term in (\ref{ent}) is the Bekenstein-Hawking
area law, while the second term is the $R^2$ generalized entropy
formula correction.

The aim of this paper is to study near-extremal $N=2$ black hole
solutions and their entropy with $R^2$-terms included. These are
non-supersymmetric solutions, the horizon is no longer $AdS_2\times
S^2$ and the attractor mechanism no longer works. The moduli of the
$CY_3$ are not fixed at the horizon in terms of the charges and the
entropy may depend on the asymptotic values of the moduli. However,
when considering the Bekenstein-Hawking entropy without $R^2$-terms, one notices that it
has the same structure as that of the extremal black holes, with the charges
being shifted \cite{kastorwin,nearextremal}
\begin{equation}
q_0 \rightarrow q_0 +\frac{1}{2}\mu h_0,~~~~~~~~p^A \rightarrow p^A
+ \frac{1}{2}\mu h^A  \ , \label{modch}
\end{equation} where $\mu$ is
the non-extremality parameter and $h_0,h^A$ correspond to the
asymptotic values of the moduli. A natural question to ask is
whether this property of the near-extremal Bekenstein-Hawking
entropy holds with the $R^2$-corrected black hole entropy
(\ref{entfor}). We construct horizon solutions for a class of near-extremal black holes with
$D_{ABC}p^Ap^Bp^C=0$, i.e. having an extremal limit with a vanishing
classical horizon area. For these black holes the entropy is the same as in the extremal case, without a shift in the charges.

Note that the $R^2$-terms considered in this paper are $F$-terms. One generally expects also $D$-term corrections, which are not taken into account here. For supersymmetric black holes, it is conjectured that such terms do not contribute to the entropy \cite{osv}.

The paper is organized as follows: In section $2$ we give a
brief review of four-dimensional $N=2$ supergravity with
$R^2$-terms. In section $3$ we derive a generalized Wald entropy
formula for non-extremal $N=2$ black holes with $R^2$-terms. In
section $4$ we present the horizon solution of near-extremal $N=2$
black hole and compute the entropy. We will first review previous
results without $R^2$-terms, and then present the new results with
$R^2$-terms.

In the paper we will use $a,b,\ldots=0,1,2,3$ to denote the tangent space
indices, corresponding to the indices $\mu,\nu,\ldots$ of the space-time
coordinates $(t,r,\phi,\theta)$. The exception are $i,j=1,2$ which are gauge $SU(2)$ indices, and $\alpha=1,2$ which is a global $SU(2)$ index.
The sign conventions for the curvature tensors follow \cite{mohauptreview}.

\section{$R^2$-Terms in $N=2$ Supergravity - A Brief Review}

We will consider $N=2$ Poincar\'e supergravity coupled to $N_V$
Abelian $N=2$ vector multiplets. $N=2$ Poincar\'e supergravity is a supersymmetric extension of
Einstein-Maxwell gravity, adding two spin $3/2$ gravitini to the
graviton and (gravi-)photon. $N=2$ Poincar\'e supergravity can
be formulated as a gauge fixed version of $N=2$ conformal
supergravity coupled to an $N=2$ Abelian vector multiplet (see
\cite{mohauptreview} for a comprehensive review).

The on-shell field content of the vector multiplet is a complex
scalar, a doublet of Weyl fermions, and a vector gauge field. We
will consider $N_V+1$ vector multiplets, and will denote by $X^I$,
$I=0\ldots N_V$, the scalars (moduli) in the vector multiplets. The
couplings of the vector multiplets are encoded in a prepotential
$F(X^I)$, which is a homogenous of second degree holomorphic
function.

The $N=2$ conformal supergravity multiplet (Weyl multiplet) is
denoted by $W^{abij}$. It consists
of gauge fields for the local symmetries: translations ($P$),
Lorentz transformations ($M$), dilatations ($D$), special conformal
transformations ($K$), $U(1)$ transformations ($A$), $SU(2)$
transformations ($V$), and supertransformations ($Q,S$). In the
theory without $R^2$-terms, the Weyl multiplet appears in the
Lagrangian through the superconformal covariantizations. In order to
get the $R^2$-terms, one adds explicit couplings to the Weyl
multiplet. This appears in the form of a background chiral multiplet, which is
equal to the square of the Weyl multiplet $W^2$. The lowest
component of the chiral multiplet is a complex scalar denoted
$\Ahat$. The prepotential $F(X^I,\Ahat)$ describes the coupling of
the vector multiplets and the chiral multiplet.

We consider a prepotential of the form
\begin{equation}
F=\frac{D_{ABC}X^AX^BX^C}{X^0}+\frac{D_AX^A}{X^0}\Ahat \ ,
\label{r2prep}
\end{equation}
where $D_{ABC},D_A$ are constants and $A,B,C=1\ldots N_V$. This
prepotential arises, for instance, from a compactification of type
IIA string theory on a Calabi-Yau three-fold. The coefficients in
the prepotential are topological data of the Calabi-Yau three-fold:
$-6D_{ABC}$ are the triple intersection numbers (symmetric in all
indices), and $-1536D_{A}$ are the second Chern class numbers. The
first term in the prepotential arises at tree-level in $\alpha'$ and
in $g_s$. The second term arises at tree-level in $\alpha'$ and is
at one-loop in $g_s$. It describes $R^2$ couplings in the
Lagrangian. In the large Calabi-Yau volume approximation
$\im(X^A/X^0)\gg1$, all other corrections are suppressed and the
prepotential consists of only these two terms. We will assume this
approximation to be valid near the horizon by an appropriate
hierarchy of charges. One introduces the
notation:
\begin{equation}
F_I\equiv\frac{\partial}{\partial X^I}F(X^I,\Ahat),~~~~~~~~
F_{\Ahat}\equiv\frac{\partial}{\partial \Ahat}F(X^I,\Ahat) \ ,
\end{equation}
and
similarly for higher order and mixed derivatives.

The bosonic part of the $N=2$ conformal supergravity Lagrangian is
\begin{eqnarray}
\label{lag} 8\pi
e^{-1}\mathcal{L}&=&-\frac{i}{2}(\bar{X}^IF_I-X^I\bar{F}_I)R+{}\nonumber\\*
&&{}+\Big(i\mathcal{D}^a\bar{X}^I\mathcal{D}_aF_I+
\frac{i}{4}F_{IJ}(F_{ab}^{-I}-\quart\bar{X}^IT_{ab}^-)
(F^{ab-J}-\quart\bar{X}^JT^{ab-})+{}\nonumber\\*
&&{}+\frac{i}{8}\bar{F}_I(F_{ab}^{-I}-\quart\bar{X}^IT_{ab}^-)T^{ab-}+
\frac{i}{32}\bar{F}T_{ab}^-T^{ab-}-\frac{i}{8}F_{IJ}Y_{ij}^IY^{ijJ}+{}\nonumber\\*
&&{}-\frac{i}{8}F_{\Ahat\Ahat}
(\widehat{B}_{ij}\widehat{B}^{ij}-2\widehat{F}_{ab}^-\widehat{F}^{ab-})+
\frac{i}{2}\widehat{F}^{ab-}F_{\Ahat
I}(F_{ab}^{-I}-\quart\bar{X}^IT_{ab}^-)+{}\nonumber\\*
&&{}-\frac{i}{4}\widehat{B}_{ij}F_{\Ahat I}Y^{ijI}+
\frac{i}{2}F_{\Ahat}\widehat{C}+\mathrm{h.c.}\Big)+{}\nonumber\\*
&&{}+i(\bar{X}^IF_I-X^I\bar{F}_I)\Big(\mathcal{D}^aV_a-\half V^aV_a-
\quart
M_{ij}\bar{M}^{ij}+D^a\Phi^i_{\phantom{i}\alpha}D_a\Phi^{\alpha}_{\phantom{\alpha}i}\Big)
\ .\nonumber\\*
\end{eqnarray}
$e\equiv\sqrt{\abs{\mathrm{det}(g_{\mu\nu})}}$ where $g_{\mu\nu}$
is the curved metric, $R$ is the Ricci scalar, $D_a$ is
the covariant derivative with respect to all superconformal
transformations, $\mathcal{D}_a$ is the covariant derivative with
respect to $P,M,D,A,V$-transformations, $F_{ab}^{-I}$ is the
anti-selfdual part of the vector field strength, $T_{ab}^-$ is an
anti-selfdual antisymmetric auxiliary field of the Weyl multiplet,
$Y_{ij}^I$ are real $SU(2)$ triplets of auxiliary
scalars of the vector multiplet, and
$V_a,M_{ij},\Phi^i_{\phantom{i}\alpha}$ are
components of a compensating nonlinear multiplet. The hatted fields are components of the
chiral multiplet $W^2$, with their bosonic parts given by
\begin{eqnarray}
\label{hatted} \theta^0~~~~~~~~\Ahat&=&T_{ab}^-T^{ab-}\nonumber\\*
\theta^2~~~~~~\widehat{B}_{ij}&=&-16R(V)_{(ij)ab}T^{ab-}\nonumber\\*
\widehat{F}^{ab-}&=&-16\mathcal{R}(M)_{cd}^{\phantom{cd}ab}T^{cd-}\nonumber\\*
\theta^4~~~~~~~~
\widehat{C}&=&64\mathcal{R}(M)_{cd}^{\phantom{cd}ab-}\mathcal{R}(M)^{cd-}_{\phantom{cd}ab}+
32R(V)_{ab\phantom{i}j}^{\phantom{ab}i-}R(V)^{abj-}_{\phantom{abj}i}-16T^{ab-}D_aD^cT_{cb}^+
\ .\nonumber\\*
\end{eqnarray}
$T_{ab}^+=\bar{T}_{ab}^-$ is the selfdual counterpart of the
auxiliary field, $R(V)_{ab\phantom{i}j}^{\phantom{ab}i}$ is the
field strength of the $SU(2)$ transformations,
$\mathcal{R}(M)_{ab}^{\phantom{ab}cd}$ is the modified Riemann
curvature and $\mathcal{R}(M)_{ab}^{\phantom{ab}cd-}$ is the
anti-selfdual projection in both pairs of indices. The bosonic part
of $\mathcal{R}(M)_{ab}^{\phantom{ab}cd}$ is given by\footnote{We
have assumed the $K$-gauge fixing which will be defined later
(\ref{gauge}).}
\begin{equation}
\label{RMS1}
\mathcal{R}(M)_{ab}^{\phantom{ab}cd}=R_{ab}^{\phantom{ab}cd}-
4f_{[a}^{\phantom{[a}[c}\delta_{b]}^{d]}+
\frac{1}{32}(T_{ab}^-T^{cd+}+T_{ab}^+T^{cd-}) \ ,
\end{equation}
where $R_{ab}^{\phantom{ab}cd}$ is the Riemann tensor, and
$f_a^{\phantom{a}c}$ is the connection of the special conformal
transformations, determined by the conformal supergravity
conventional constraints, with the bosonic
part\footnotemark[\value{footnote}]:
\begin{equation}
f_a^{\phantom{a}c}=\half
R_a^{\phantom{a}c}-\quart(D+\frac{1}{3}R)\delta_a^c+\frac{1}{2}{}^{\star}R(A)_a^{\phantom{a}c}+
\frac{1}{32}T_{ab}^-T^{cb+} \ ,
\end{equation}
where $R_a^{\phantom{a}c}$ is the Ricci tensor, $D$ is an
auxiliary real scalar field of the Weyl multiplet, and
${}^{\star}R(A)^a_{\phantom{a}b}$ is the Hodge dual of the field
strength of the $U(1)$ transformations. Note that the $T^2$-terms
in $\mathcal{R}(M)_{ab}^{\phantom{ab}cd}$ cancel exactly the $T^2$
contribution from $f_a^{\phantom{a}c}$.

The auxiliary field $D$ is constrained by a constraint on the
nonlinear multiplet:
\begin{equation}
\mathcal{D}^aV_a-D-\frac{1}{3}R-\half V^aV_a-\quart
M_{ij}\bar{M}^{ij}+D^a\Phi^i_{\phantom{i}\alpha}D_a\Phi^{\alpha}_{\phantom{\alpha}i}=0
\ , \label{nl}
\end{equation}
where we have assumed a bosonic solution.

In order to obtain Poincar\'e supergravity one gauge fixes the
bosonic fields (in addition there is a gauge fixing of fermionic
fields):
\begin{eqnarray}
K\mathrm{-gauge}:&&b_a=0\nonumber\\*
D\mathrm{-gauge}:&&i(\bar{X}^IF_I-X^I\bar{F}_I)=1\nonumber\\*
A\mathrm{-gauge}:&&X^0=\bar{X}^0>0\nonumber\\*
V\mathrm{-gauge}:&&\Phi^i_{\phantom{i}\alpha}=\delta_{\alpha}^i \ ,
\label{gauge}
\end{eqnarray}
where $b_a$ is the connection of the dilatations.

\section{Entropy Formula with $R^2$-Terms}

With the addition of $R^2$-terms to the Lagrangian, the
Bekenstein-Hawking entropy formula is no longer valid. A
generalization of the area law has been derived by Wald \cite{wald}.
The Bekenstein-Hawking area is recovered when taking the
Einstein-Hilbert Lagrangian. We work with the $N=2$ supergravity
Lagrangian, which does not depend on derivatives of the Riemann
tensor, and we further assume the black holes to be static and
spherically symmetric. The generalized entropy formula in this case
is
\begin{equation}
\label{wald} S=2\pi
A\ve_{ab}\ve^{cd}\frac{\partial(e^{-1}\mathcal{L})}{\partial
R_{ab}^{\phantom{ab}cd}} \ ,
\end{equation}
where $A$ is the (modified) area of the horizon,
$\ve_{01}=-\ve_{10}=1$, $\mathcal{L}$ is the Lagrangian density, and
the expression is evaluated on the event horizon. In the derivative,
we treat the Riemann tensor and the metric as being independent and take
into account the supergravity constraints on the fields. Our
derivation is similar to that of \cite{r2entropy,mohauptreview}.

For the Lagrangian (\ref{lag}) we get
\begin{eqnarray}
\label{dldr} \frac{\partial(e^{-1}\mathcal{L})}{\partial
R_{ab}^{\phantom{ab}cd}}&=&-\frac{1}{16\pi}\delta_c^a\delta_d^{b}+{}\nonumber\\*
&&{}-\frac{1}{8\pi}\im\left(F_{\Ahat I}
(F_{ef}^{-I}-\quart\bar{X}^IT_{ef}^-)\frac{\partial\widehat{F}^{ef-}}{\partial
R_{ab}^{\phantom{ab}cd}}+F_{\Ahat\Ahat}\widehat{F}_{ef}^-\frac{\partial\widehat{F}^{ef-}}{\partial
R_{ab}^{\phantom{ab}cd}}+F_{\Ahat}\frac{\partial\widehat{C}}{\partial
R_{ab}^{\phantom{ab}cd}}\right) \ .\nonumber\\*
\end{eqnarray}

This expression can be simplified to\footnote{Using the identity
for anti-selfdual tensors:
$R^{mn-}_{\phantom{mn}pq}R_{mn}^{\phantom{mn}pq-}=R^{mn-}_{\phantom{mn}pq}R_{mn}^{\phantom{mn}pq}$.}:
\begin{eqnarray}
\frac{\partial(e^{-1}\mathcal{L})}{\partial
R_{ab}^{\phantom{ab}cd}}&=&-\frac{1}{16\pi}\delta_c^a\delta_d^b+{}\nonumber\\*
&&{}+\frac{1}{\pi}\im\Bigg[\Big(2F_{\Ahat
I}(F_{pq}^{-I}-\quart\bar{X}^IT_{pq}^-)T^{mn-}-
32F_{\Ahat\Ahat}\mathcal{R}(M)^{xy}_{\phantom{xy}pq}T_{xy}^-T^{mn-}+{}\nonumber\\*
&&\qquad\qquad-16F_{\Ahat}\mathcal{R}(M)^{mn-}_{\phantom{mn}pq}\Big)
\frac{\partial\mathcal{R}(M)_{mn}^{\phantom{mn}pq}}{\partial
R_{ab}^{\phantom{ab}cd}}-F_{\Ahat}T^{an-}T_{cn}^+\delta_d^b\Bigg] \
. \nonumber\\*
\end{eqnarray}
The last term is the same as in the
supersymmetric case\footnote{Using that
$D_aD^cT_{cb}^+=\mathcal{D}_a\mathcal{D}^cT_{cb}^+-f_a^{\phantom{a}c}T_{cb}^+$,
for a bosonic solution. The covariant derivative string may be
expanded as
$\mathcal{D}_a\mathcal{D}^c=\half\{\mathcal{D}_a,\mathcal{D}^c\}+\half[\mathcal{D}_a,\mathcal{D}^c]$.
Only the anticommutator part is dependent on the Riemann tensor,
however its contribution vanishes due to the identity for
(anti-)selfdual tensors:
$T^{ab-}T^{c+}_{\phantom{c}b}=T^{cb-}T^{a+}_{\phantom{a}b}$.}.

We have the relation:
\begin{equation}
\label{RMS2}
\mathcal{R}(M)^{mn}_{\phantom{mn}pq}=C^{mn}_{\phantom{mn}pq}+D\delta_{[p}^{[m}\delta_{q]}^{n]}-
2\delta_{[p}^{[m}{}^{\star}R(A)^{n]}_{\phantom{n]}q]} \ ,
\end{equation}
where $C^{ab}_{\phantom{ab}cd}$ is the Weyl tensor. In addition,
from the definition of $\mathcal{R}(M)_{ab}^{\phantom{ab}cd}$ (\ref{RMS1}) we
get:
\begin{equation}
\frac{\partial\mathcal{R}(M)_{mn}^{\phantom{mn}pq}}{\partial
R_{ab}^{\phantom{ab}cd}}\sim\delta_m^a\delta_n^b\delta_c^p\delta_d^q-
2\delta_{[m}^{[p}\delta_{n]}^a\delta_c^{q]}\delta_d^b \ ,
\end{equation}
where the RHS must be constrained to have the same symmetries as the
LHS, and we used $D=-\frac{1}{3}R+\ldots$ due to the nonlinear
multiplet constraint (\ref{nl}).

Substituting all expressions, we obtain
the generalized entropy formula for the non-extremal $R^2$ case:
\begin{equation}
\label{genentropy} S=\quart
A-4A\cdot\im\Big(F_{\Ahat}(\abs{T_{01}^-}^2+16C_{0101}+16D)\Big) \ ,
\end{equation}
where we have used spherical symmetry, everything is evaluated on
the event horizon, and
$\Ahat=-4(T_{01}^-)^2$.
This formula differs from the extremal $R^2$ case
by the $C_{0101}$ and $D$ terms, where one also had
\begin{equation}
\Ahat= -256\pi A^{-1} \ .
\end{equation}
Note that as in
the extremal $R^2$ case, the entropy does not depend on the higher
order derivatives $F_{\Ahat I},F_{\Ahat\Ahat}$.

\section{Near-Extremal $N=2$ Black Holes}

\subsection{Near-Extremal $N=2$ Black Holes without
$R^2$-Terms}

We will start by discussing non-extremal black holes in $N=2$
supergravity without $R^2$-terms \cite{kastorwin, nearextremal}.
The metric is given by
\begin{equation}
ds^2=-e^{-2U(r)}f(r)dt^2+e^{2U(r)}(f(r)^{-1}dr^2+r^2d\Omega^2) \ ,
\label{metric}
\end{equation}
where $d\Omega^2=\sin^2{\theta}d\phi^2+d\theta^2$, and
\begin{equation}
f(r)=1-\frac{\mu}{r} \ , \label{nonext}
\end{equation}
and $\mu\geq0$ is a non-extremality parameter. The background is non-supersymmetric, with $\mu$ parameterizing the difference between the ADM mass and the BPS mass.

The event horizon is located at $r=\mu$ and the inner horizon at
$r=0$. Unlike the extremal black holes, the event
horizon geometry is not $AdS_2{\times}S^2$.

Consider the prepotential:
\begin{equation}
\label{prep} F=\frac{D_{ABC}X^AX^BX^C}{X^0} \ ,
\end{equation}
and the ansatz
\begin{equation}
\label{ukansatz} e^{2U(r)}=e^{-K} \ ,
\end{equation}
where the K\"ahler potential $K$ is
\begin{equation}
e^{-K}=i\Big(\bar{X}^I(\bar{z})F_I(z)-X^I(z)\bar{F}_I(\bar{z})\Big) \ .
\label{kahler}
\end{equation}
$F_I(z)=F_I(X(z))$ and $X^I(z)$ are
related to the $X^I$, by
\begin{equation}
X^I=e^{\half K}X^I(z) \ .
\end{equation}

Consider black holes with one electric charge $q_0$ and $p^A$ ($A=1\ldots N_V$) magnetic charges. One introduces the boost parameters
$\gamma^A,\gamma_0$, related to the charges by
\begin{eqnarray}
p^A&=&h^A\mu\sinh\gamma^A\cosh\gamma^A\qquad\textrm{(no
summation)}\nonumber\\* q_0&=&h_0\mu\sinh\gamma_0\cosh\gamma_0 \ ,
\end{eqnarray}
where $h^A,h_0$ are constants\footnote{These parameters are
constrained by the asymptotic flatness condition:
$e^{2U(\infty)}=\abs{h^IF_I(\infty)-h_IX^I(\infty)}^2=1$.} that
determine the moduli at infinity. Note that for fixed charges and
non-extremality parameter, a choice of $(\gamma^A,\gamma_0)$ is
equivalent to a choice of $(h^A,h_0)$. The extremal case is
recovered in the limit
$\mu\rightarrow0;(\gamma^A,\gamma_0)\rightarrow\infty$, with the
 charges held fixed.

Introduce the modified charges
\begin{eqnarray}
\label{pqtilde}
\tilde{p}^A&\equiv&h^A\mu\sinh^2\gamma^A=\alpha^Ap^A\qquad\textrm{(no
summation)}\nonumber\\*
\tilde{q}_0&\equiv&h_0\mu\sinh^2\gamma_0=\alpha_0q_0 \ ,
\end{eqnarray}
where $\alpha^A\equiv\tanh\gamma^A,\alpha_0\equiv\tanh\gamma_0$.
In the extremal case $(\alpha^A,\alpha_0)\rightarrow1$.

In the extremal supersymmetric case, the vanishing of the gaugino
variations under $N=1$ supertransformations, implies generalized
stabilization equations, also called the supersymmetric attractor
mechanism \cite{genstabeq}. These equations determine the values of
the moduli on the horizon in terms of the electric and magnetic
charges. In the non-extremal case the gaugino variations do not
vanish. Consider an ansatz similar to the supersymmetric
stabilization equations of the form
\begin{eqnarray}
i(X^I(z)-\bar{X}^I(\bar{z}))&=&\tilde{H}^I\nonumber\\*
i(F_I(z)-\bar{F}_I(\bar{z}))&=&\tilde{H}_I \ , \label{modified}
\end{eqnarray}
where $\tilde{H}^I,\tilde{H}_I$ are harmonic functions
\begin{eqnarray}
\tilde{H}^I&=&h^I+\frac{\tilde{p}^I}{r}\nonumber\\*
\tilde{H}_I&=&h_I+\frac{\tilde{q}_I}{r} \ .
\end{eqnarray}
These equations do not exhibit an attractor behavior, since the
moduli on the event horizon at $r=\mu$ depend on the moduli at
infinity.

The ansatz solves the field equations for equal parameters
$\gamma^A$ ($A=1\ldots N_V$). One can relax some of the
conditions on the $\gamma^A$'s by restricting the prepotential to
specific choices $D_{ABC}$. For instance, if only one of the $D_{ABC}$'s (up to permutations) is
nonzero, all $\gamma^A$'s may be chosen independently.\footnote{Alternatively, \cite{nearextremal} suggests a method for finding
near-extremal solutions with no restrictions but only in the
near-extremal regime.}

For the case of either equal $\gamma^A$ ($A=1,2,3$) or only $D_{123}\neq0$, the auxiliary field $T_{ab}^-$ takes the form
\begin{equation}
T_{01}^-=iT_{23}^-=\left(\frac{k_0}{\alpha_0(r+k_0)}+\frac{k^1}{\alpha^1(r+k^1)}+\frac{k^2}{\alpha^2(r+k^2)}+\frac{k^3}{\alpha^3(r+k^3)}\right)\frac{1}{r}e^{-U(r)}
\ , \label{aux}
\end{equation}
where
\begin{eqnarray}
k^A&\equiv&\mu\sinh^2\gamma^A=\frac{\mu(\alpha^A)^2}{1-(\alpha^A)^2}\qquad\textrm{(no
summation)}\nonumber\\*
k_0&\equiv&\mu\sinh^2\gamma_0=\frac{\mu(\alpha_0)^2}{1-(\alpha_0)^2}
\ .
\end{eqnarray}

Solving the stabilization equations, one obtains the moduli on the
horizon in terms of the charges and the moduli at infinity. The
Bekenstein-Hawking entropy takes the form
\begin{equation}
\label{qspspsps} S=\quart A=2\pi\sqrt{\qs_0D_{ABC}\ps^A\ps^B\ps^C} \
,
\end{equation}
where
\begin{eqnarray}
\ps^A&\equiv&\frac{p^A}{\alpha^A}=h^A\mu\cosh^2\gamma^A\qquad\textrm{(no
summation)}\nonumber\\*
\qs_0&\equiv&\frac{q_0}{\alpha_0}=h_0\mu\cosh^2\gamma_0 \ .
\end{eqnarray}
This has the same form as the extremal entropy, with the charges
$(q_0,p^A)$ replaced by $(\qs_0,\ps^A)$. Note that, unlike the
extremal case, the entropy depends on the values of the moduli at
infinity. In addition, the non-extremal entropy has a different
functional dependence on the original charges since the parameters
$\alpha^A,\alpha_0$ themselves depend on the charges for given asymptotic moduli.

The near-extremal black holes are described by adding to the
extremal black holes the leading terms in $\mu$, while holding the
physical charges fixed. One gets
\begin{eqnarray}
\ps^A&=&p^A+\half h^A\mu+O(\mu^2)\nonumber\\* \qs_0&=&q_0+\half
h_0\mu+O(\mu^2) \ .
\end{eqnarray}
We see that the near-extremal Bekenstein-Hawking entropy formula has
the same structure as the extremal entropy with a modification of
the charges depending on the non-extremality parameter $\mu$ and the
asymptotic values of the moduli $h^A$. In the next section we will
construct a class of horizon solutions, where this structure holds
with $R^2$-terms, as in (\ref{entfor}) and (\ref{modch}).

\subsection{Near-Extremal $N=2$ Black Holes with
$R^2$-Terms}

We would like to get an explicit expression for the entropy for the
near-extremal black holes with $R^2$-terms (\ref{genentropy}), as a
function of the charges and the moduli at infinity. Consider black
holes with one electric charge $q_0$ and $p^A$ ($A=1,2,3$) magnetic
charges.

Let us introduce the dual field strength:
\begin{equation}
G_{abI}^-=2i\frac{\partial(e^{-1}\mathcal{L})}{\partial F^{ab-I}}=F_{IJ}F_{ab}^{-J}+\quart(\bar{F}_I-F_{IJ}\bar{X}^J)T_{ab}^-+F_{\Ahat I}\widehat{F}_{ab}^- \ ,
\end{equation}
where we have considered only bosonic terms. Due to spherical symmetry we have
\begin{eqnarray}
F_{23}^{-I}&=&-iF_{01}^{-I}\nonumber\\*
G_{23I}^-&=&-iG_{01I}^- \ .
\end{eqnarray}
The field strengths $F_{01}^{-I}$ may be extracted from the following equations:
\begin{equation}
2(\im F_{IJ})F_{01}^{-J}=G_{23I}-\bar{F}_{IJ}F_{23}^J+\half\im\Big(\left(F_I+F_{IJ}\bar{X}^J-64F_{\Ahat I}(2C_{0101}-D)\right)T_{01}^-\Big) \ ,
\end{equation}
where we used spherical symmetry and (\ref{RMS2}). The magnetic parts of the field strengths are obtained from Bianchi identities, which for a static spherically symmetric metric give:
\begin{eqnarray}
F_{23}^I&=&\frac{1}{r^2}e^{-2U(r)}p^I\nonumber\\*
G_{23I}&=&\frac{1}{r^2}e^{-2U(r)}q_I \ ,
\end{eqnarray}
where we used $g_{\theta\theta}=g_{\phi\phi}/sin^2\theta=r^2e^{2U(r)}$.
For our choice of charges, and the complex-valued form of our ansatz (\ref{r2nesol}) we get
\begin{eqnarray}
\label{fsol}
F_{01}^{-0}&=&\frac{1}{2F_{00}}\Big(iG_{230}-iF_{0A}F_{23}^A+\half\left(F_0+F_{0I}\bar{X}^I-64F_{\Ahat0}(2C_{0101}-D)\right)T_{01}^-\Big)\nonumber\\*
F_{01}^{-A}&=&\frac{i}{2}F_{23}^A \ .
\end{eqnarray}

The auxiliary field $D$ may be determined by the constraint on the
nonlinear multiplet (\ref{nl}):
\begin{equation}
D=\mathcal{D}^aV_a-\frac{1}{3}R-\half V^aV_a-\quart
M_{ij}\bar{M}^{ij}+D^a\Phi^i_{\phantom{i}\alpha}D_a\Phi^{\alpha}_{\phantom{\alpha}i}
\ .
\end{equation}
This means that we must make the above substitution also for $D$ appearing in the hatted fields (\ref{hatted}) in the
Lagrangian (\ref{lag}). We assume
\begin{eqnarray}
V_a&=&0\nonumber\\*
M_{ij}&=&0\nonumber\\*
\Phi^i_{\phantom{i}\alpha}&=&\delta_{\alpha}^i \ ,
\end{eqnarray}
where the later equation is the $V$-gauge (\ref{gauge}).
The equation of motion for $V_a$ must be shown to be satisfied. The equations of motion for $M_{ij}$ and $\Phi^i_{\phantom{i}\alpha}$ are trivially satisfied by the above assumption, where for the latter we assume a vanishing $SU(2)$ connection as we shall consider later.
We therefore remain with the constraint:
\begin{equation}
D=-\frac{1}{3}R \ .
\end{equation}

The area of the horizon $A$, the Weyl tensor $C_{0101}$, and the
Ricci scalar $R$ are all calculated from the metric. It remains to
find solutions for the metric, the moduli $X^I(z)$, and the
auxiliary field $T_{01}^-$. In addition, for solving the equations
of motion, we will need solutions for the $U(1)$ connection $A_a$
and the $SU(2)$ connection
$\mathcal{V}_{a\phantom{i}j}^{\phantom{a}i}$, which in the
supersymmetric case could be taken as zero. We will make an ansatz
for the solution on the horizon, which is an extension of both the
extremal case with $R^2$-terms (see \cite{mohauptreview}) and the
non-extremal case without $R^2$-terms. One may consider the ansatz
of the non-extremal case for the metric (\ref{metric}),
(\ref{ukansatz}), (\ref{nonext}), the modified stabilization
equations (\ref{modified}) which give the moduli, and the auxiliary
field (\ref{aux}), with the $R^2$ prepotential (\ref{r2prep}).
However this proves to be insufficient, and since we will consider a
near-extremal solution, we introduce linear $\mu$-corrections to the
fields.

Our ansatz is
\begin{eqnarray}
\label{r2nesol}
F&=&\frac{D_{ABC}X^AX^BX^C}{X^0}+\frac{D_AX^A}{X^0}\Ahat\nonumber\\*
ds^2&=&-e^{-2U(r)}f(r)dt^2+e^{2U(r)}(f(r)^{-1}dr^2+r^2d\Omega^2)\nonumber\\*
e^{2U(r)}&=&e^{-K}(1+\mu\beta_U)\nonumber\\*
f(r)&=&\left(1-\frac{\mu}{r}\right)(1+\mu\beta_f)\nonumber\\*
X^A(z)&=&-\frac{i}{2}x^A(1+\mu\beta_A)\nonumber\\*
X^0(z)&=&\half\sqrt{\frac{D_{ABC}x^Ax^Bx^C-4D_Ax^A\Ahat(z)}{x_0}}(1+\mu\beta_0)\nonumber\\*
T_{01}^-&=&iT_{23}^-=\left(\frac{k_0}{\alpha_0(r+k_0)}+\frac{k^1}{\alpha^1(r+k^1)}+\frac{k^2}{\alpha^2(r+k^2)}+\frac{k^3}{\alpha^3(r+k^3)}\right)\frac{1}{r}e^{\half
K}(1+\mu\beta_T)\ ,\nonumber\\*
\end{eqnarray}
where
\begin{eqnarray}
x^A&\equiv&\frac{\alpha^Ap^A}{k^A}+\frac{\alpha^Ap^A}{r}\qquad\textrm{(no
summation)}\nonumber\\*
x_0&\equiv&\frac{\alpha_0q_0}{k_0}+\frac{\alpha_0q_0}{r} \ ,
\end{eqnarray}
and
\begin{eqnarray}
\Ahat(z)&=&e^{-K}\Ahat=-4e^{-K}(T_{01}^-)^2=\nonumber\\*
&=&-4\left(\frac{k_0}{\alpha_0(r+k_0)}+\frac{k^1}{\alpha^1(r+k^1)}+\frac{k^2}{\alpha^2(r+k^2)}+\frac{k^3}{\alpha^3(r+k^3)}\right)^2\frac{1}{r^2}(1+\mu\beta_T)^2
\ .\nonumber\\*
\end{eqnarray}
$\beta_U,\beta_f,\beta_1,\beta_2,\beta_3,\beta_0,\beta_T$ are constants satisfying $\mu\abs{\beta}\ll1$, and
$e^{-K}$ also contains $\beta$'s. Note that besides the explicit
$\mu$-corrections above, some of the fields will also have implicit
$\mu$ dependence via $e^{-K}$ and $A(z)$.

In addition we assume
\begin{eqnarray}
A_a&=&0\nonumber\\* \mathcal{V}_{a\phantom{i}j}^{\phantom{a}i}&=&0 \
.
\end{eqnarray}
The equation of motion for the $SU(2)$ connection is
always satisfied by the vanishing $SU(2)$ connection, for a
bosonic background and with our choice of $V$-gauge (also assuming
no hyper-multiplet scalars). This is because the $SU(2)$ connection
and its derivatives, appear then in the Lagrangian (\ref{lag}) always in at least
a quadratic form.\footnote{More generally, the $SU(2)$ field equations are automatically satisfied since the solution is a singlet under the $SU(2)$ symmetry.}
The vanishing of the $SU(2)$ connection implies also $Y_{ij}^I=0$
\cite{yeom}.

For our ansatz to constitute a solution, it must satisfy the
equations of motion on the horizon for the metric, the moduli
$X^I(z)$, the auxiliary field $T_{01}^-$, the $U(1)$
connection $A_a$, and the nonlinear multiplet field $V_a$. In general the above ansatz is not a solution
to the equations of motion. However, we have found that it may
constitute a near-extremal horizon solution if we require equal
boost parameters and $D_{ABC}p^Ap^Bp^C=0$. The latter condition on
the charges implies the vanishing of the classical horizon area in
the extremal limit.\footnote{Note added: In a later paper we construct non-extremal $R^2$ solutions in all space with $D_{ABC}p^Ap^Bp^C\neq0$, by using the large charge approximation. The entropy of these solutions exhibits the same charge replacement property as seen in the $R$-level solutions. [arXiv:0902.3799]}

In the near-extremal regime we linearize the algebraic equations of
motion (after substituting the ansatz) in the small expansion
parameter $\mu\ll(2k^A,2k_0)$. Recall that the boost parameters $\alpha^A,\alpha_0$ depend on $\mu$ with constant $k^A,k_0$:
\begin{eqnarray}
\alpha^A&=&\sqrt{\frac{k^A}{k^A+\mu}}\qquad\textrm{(no
summation)}\nonumber\\* \alpha_0&=&\sqrt{\frac{k_0}{k_0+\mu}} \ .
\end{eqnarray}
The expansion must be done after taking the
horizon limit $r\rightarrow\mu$, since for a small but finite $\mu$
we want to have two topologically distinct horizons in the metric function.
Next, we choose equal boost parameters:
$\alpha_0=\alpha^1=\alpha^2=\alpha^3$. Without $R^2$-terms, this
would be the non-extremal version of the double-extremal black hole.
Denote: $k\equiv k_0=k^1=k^2=k^3$. Finally, $D_{ABC}p^Ap^Bp^C=0$
would imply a vanishing of the classical horizon area for the
extremal $R$-level case (i.e. without $R^2$-terms) \cite{smallbh}.

Under these three restrictions our ansatz (\ref{r2nesol}) solves the
field equations, with the $\beta$'s for some simplified cases given
in appendix $A$. In appendix $B$ we comment on the derivation of the metric
field equations.

For the above ansatz, the area of the horizon, the Ricci scalar and
the Weyl tensor on the horizon read
\begin{eqnarray}
A&=&4{\pi}\mu^2e^{-K(r=\mu)}\nonumber\\*
R&=&\frac{\mu\beta_f}{8\sqrt{q_0D_Ap^A}}+O(\mu^2)\nonumber\\*
C_{0101}&=&-\frac{\mu\beta_f}{48\sqrt{q_0D_Ap^A}}+O(\mu^2) \ .
\end{eqnarray}

Substituting the solution in the generalized entropy formula for the
near-extremal $R^2$ case (\ref{genentropy}) yields the explicit
entropy:
\begin{equation}
S=32\pi\sqrt{q_0D_Ap^A}+O(\mu^2),
\end{equation}
where we must choose the signs of the charges such that the result
is real. It turns out that the linear $\mu$-terms which appear in the Bekenstein-Hawking entropy and in the Wald correction to the entropy, exactly cancel. Thus to this order of approximation, the entropy does not depend on $\mu$, nor on the asymptotic moduli at infinity, and is the same as in the extremal case. This does not exhibit the same shift of charges as in the transition from the $R$-level extremal entropy to the near-extremal entropy. It would be
interesting to compare the obtained expression for the entropy, to a
corresponding microscopic statistical entropy, which is currently
unknown.

We have considered only the tree-level $\alpha'$ $F$-terms. For this we require that the large volume
approximation is valid near the horizon, by imposing:
$\abs{q_0}\gg\abs{p^3}$. We may further take the magnetic charges to be large, in order to damp out any $R^4$ or higher $D$-term corrections.
However, we cannot rule out other contributions to the field
solutions and entropy coming from $D$-terms at the $R^2$-level.

The Hawking temperature for our static spherically symmetric black
hole is given by
\begin{equation}
T=-\frac{\partial_r
g_{tt}}{4\pi\sqrt{-g_{tt}g_{rr}}}\Bigg|_{horizon}=\frac{\mu}{64\pi\sqrt{q_0D_Ap^A}}+O(\mu^3)
\ .
\end{equation}
Note that the quadratic term in $\mu$ vanishes. Also, since the entropy has a vanishing linear term in $\mu$, the first law of thermodynamics implies that the mass has vanishing linear and quadratic terms in $\mu$.\footnote{$
T=\frac{\partial M}{\partial S}=\frac{\partial M(q_I,p^I,h_I,h^I,\mu)}{\partial\mu}\left(\frac{\partial S(q_I,p^I,h_I,h^I,\mu)}{\partial\mu}\right)^{-1}$, where $M$ is the mass, $q_I,p^I,h_I,h^I$ are held fixed, and assuming $\partial S/\partial\mu\neq0$.}

Since the solutions have been constructed only on the horizon, and
without the supersymmetry property, one still needs to analyze
whether an interpolating solution exists which smoothly connects the
horizon to asymptotically flat space. This is a prerequisite for the
existence of a corresponding black hole and for the validity of the
Wald entropy formula (\ref{wald}).

\section*{Acknowledgements}

We would like to thank I. Adam, B. de Wit, D. Gl\"{u}ck, N.
Itzhaki and H. Nieder for valuable discussions.

\appendix
\section{Solutions of the Field Equations}
Following are the explicit solutions for the $\beta$'s of
(\ref{r2nesol}) which satisfy the equations of motion, under the
discussed restrictions. The free real parameter $\beta_U$, represents a gauge freedom that should be fixed
by the normalization of the interpolating solution at infinity. Our calculations
were done using Maple with GRTensor.

(i) For the case
$D_{113}=D_{133}=D_{223}=D_{233}=D_{123}=D_1=D_2=0$:
\begin{eqnarray}
\beta_f&=&-\frac{3D_{333}p^3p^3}{256kD_3}\nonumber\\*
2\beta_0&=&\beta_f\nonumber\\*
2\beta_1=2\beta_2&=&\beta_f-\beta_U\nonumber\\*
2\beta_3&=&\beta_f-\frac{1}{k}-\beta_U\nonumber\\*
2\beta_T&=&\frac{1}{k}-\beta_U \ .
\end{eqnarray}

(ii) For the case
$D_{112}=D_{122}=D_{223}=D_{233}=D_{222}=D_{123}=D_2=0$:
\begin{eqnarray}
\beta_f&=&\frac{3X^2}{256kY}\nonumber\\*
2\beta_0&=&\beta_f\nonumber\\*
2\beta_1&=&\beta_f-\frac{D_3p^3X}{kY}-\frac{1}{k}-\beta_U\nonumber\\*
2\beta_2&=&\mathrm{unconstrained}\nonumber\\*
2\beta_3&=&\beta_f+\frac{D_1p^1X}{kY}-\frac{1}{k}-\beta_U\nonumber\\*
2\beta_T&=&\frac{1}{k}-\beta_U \ ,
\end{eqnarray}
where
\begin{eqnarray}
X&=&\half\left(D_{333}p^3p^3p^3+D_{133}p^1p^3p^3-D_{113}p^1p^1p^3-D_{111}p^1p^1p^1\right)\nonumber\\*
Y&=&\left(D_1D_{333}p^3p^3+D_1D_{133}p^1p^3+D_3D_{113}p^1p^3+D_3D_{111}p^1p^1\right)p^1p^3
\ .\nonumber\\*
\end{eqnarray}
Note that here it is assumed that $Y\neq0$ and
\begin{equation}
D_{333}p^3p^3p^3+3D_{133}p^1p^3p^3+3D_{113}p^1p^1p^3+D_{111}p^1p^1p^1=0
\ .
\end{equation}

(iii) For the case
$D_{112}=D_{122}=D_{113}=D_{133}=D_{223}=D_{233}=D_{111}=D_{222}=D_{333}=1$,
$D_{123}=-\frac{7}{2}$, and $p^1=p^2=p^3$:
\begin{eqnarray}
\beta_f&=&0\nonumber\\* 2\beta_0&=&0\nonumber\\*
2\beta_1=2\beta_2=2\beta_3&=&-\frac{1}{k}-\beta_U\nonumber\\*
2\beta_T&=&\frac{1}{k}-\beta_U \ .
\end{eqnarray}

\section{Derivation of the Metric Field Equations}
In order to simplify the derivation of the equations of motion, we
write the Lagrangian in a form which is explicit in the scalar
degrees of freedom. There are some subtleties regarding the degrees
of freedom of the metric. Here we will identify these degrees of
freedom and how they should be accounted for in the computation.

Let $\mathcal{L(\psi,\partial_\mu \psi,\partial_\mu\partial_\nu
\psi)}$ be a Lagrangian density depending on the scalar field $\psi$
and its first and second space-time derivatives. The equation of
motion for $\psi$ is given by the Euler-Lagrange equation:
\begin{equation}
\frac{\partial\mathcal{L}}{\partial\psi}-\partial_\mu\left(\frac{\partial\mathcal{L}}{\partial(\partial_\mu\psi)}\right)
+\partial_\mu\partial_\nu\left(\frac{\partial\mathcal{L}}{\partial(\partial_\mu\partial_\nu\psi)}\right)=0
\ .
\end{equation}
In our case, the action contains curvature tensors which are built
from second order derivatives. Thus we need to take the full second
order variation. Alternatively, one may integrate the action by
parts, and take the usual first order variation.

We assume a static and spherically symmetric metric. A general form
of such a metric is
\begin{equation}
\label{u1u2u3metric}
ds^2=-e^{-2U_1(r)}dt^2+e^{2U_2(r)}dr^2+e^{2U_3(r)}r^2d\Omega^2 \ .
\end{equation}
Correspondingly, we will get three equations of motion for
$U_1(r),U_2(r),U_3(r)$. Any metric has only two real degrees of
freedom: $16$ $-$ $6$ (symmetric components) - $4$ (Bianchi
identities) - $4$ (coordinate redefinitions) $=$ $2$. So our static
and spherically symmetric metric really contains only two
independent $r$-function degrees of freedom. Thus one of the three
equations of motion will be redundant.

We will explain why the above policy is nevertheless advantageous.
One may set e.g. $U_2(r)=U_1(r)$ by a redefinition of the
$r$-coordinate. This choice of ``gauge'' may result in a trivial
equation of motion for $U_1(r)$, leaving us with only one
independent equation of motion. The missing equation of motion has
to be obtained from the requirement that the action is invariant
under the choice of gauge. I.e. the variation of the
non-gauged-fixed action with respect to $U_2(r)$ must vanish. This
is also known as a Hamiltonian constraint. However, this is just the
original equation of motion for $U_2(r)$ that we threw away by the
gauge fixing. Thus we will simply retain all three degrees of
freedom in the metric, which will give two independent equations of motion after gauge fixing.

One may now be concerned about other gauge fixings implicit in the
choice of coordinates of (\ref{u1u2u3metric}), e.g. vanishing
off-diagonal components or
$g_{\theta\theta}=g_{\phi\phi}/sin^2\theta$. However, our metric is
the ``maximally general'' metric preserving the assumed isometries
of the solution, namely staticity and spherical symmetry
\cite{barakol}. For such a solution, the equations of motion
corresponding to the trivial metric components would be
automatically satisfied and would not yield new constraints.

In order to be consistent with the notation of our solution
(\ref{r2nesol}), we will actually use the metric:
\begin{equation}
ds^2=-e^{-2U_1(r)}f(r)dt^2+e^{2U_2(r)}f(r)^{-1}dr^2+e^{2U_3(r)}r^2d\Omega^2
\ ,
\end{equation}
where $f(r)$ is given. The solution to the equations of motion is
given by
\begin{equation}
U_1(r)=U_2(r)=U_3(r)=U(r) \ ,
\end{equation}
where $U(r)$ is given. When deriving the equations of motion, we
must retain the separate degrees of freedom of the metric.

The fields $F_{ab}^{-I},T_{ab}^-$ in our solution, are given as the
anti-selfdual parts written with tangent space indices. In this
form, these fields contain metric components, while the
metric-independent fields are $F_{\mu\nu}^{I},T_{\mu\nu}$. Let us
denote by $F_{01}^{-I}(r),T_{01}^{-I}(r)$ the $(0,1)$ components of
these fields as given in our solution (\ref{fsol}), (\ref{r2nesol}),
before we explicitly introduced the separate metric degrees of
freedom. When these fields appear in the Lagrangian explicitly
(including via the hatted fields (\ref{hatted})), they should be
rewritten as
\begin{eqnarray}
F_{01}^{-A}&=&iF_{23}^{-A}=e^{2U(r)-2U_3(r)}F_{01}^{-A}(r)\nonumber\\*
F_{01}^{-0}&=&iF_{23}^{-0}=e^{U_1(r)-U_2(r)}F_{01}^{-0}(r)\nonumber\\*
T_{01}^-&=&iT_{23}^-=e^{U_1(r)-U_2(r)}T_{01}^-(r) \ .
\end{eqnarray}
Alternatively, one may work with the $F_{\mu\nu}^{I},T_{\mu\nu}$
form and put appropriate projection operators in the Lagrangian.


\newpage

\begin{thebibliography}{99}

\bibitem{mohauptreview}
T. Mohaupt, ``Black hole entropy, special geometry and strings'',
Fortsch. Phys. 49 (2001) 3, arXiv:hep-th/0007195.

\bibitem{recentreviews}
T. Mohaupt, ``Strings, Higher curvature corrections, and black
holes'', arXiv:hep-th/0512048; B. Pioline, ``Lectures on black
holes, topological strings and quantum attractors'', Class. Quant.
Grav. 23 (2006) S981, arXiv:hep-th/0607227; T. Mohaupt, ``Supersymmetric black
holes in string theory'', Fortsch. Phys. 55 (2007) 519, arXiv:hep-th/0703035; T. Mohaupt, ``Special geometry, black holes and Euclidean supersymmetry'', arXiv:hep-th/0703037;
B. de Wit, ``BPS black holes'', Nucl. Phys. Proc. Suppl. 171 (2007) 16, arXiv:0704.1452 [hep-th]; A. Sen, ``Black hole entropy
function, attractors and precision counting of microstates'', Gen. Rel. Grav. 40 (2008) 2249, arXiv:0708.1270 [hep-th].

\bibitem{nonsusy}
P. K. Tripathy and S. P. Trivedi, ``Non-supersymmetric attractors in string theory'', JHEP 0603 (2006) 022, arXiv:hep-th/0511117; R. Kallosh, N. Sivanandam and M. Soroush, ``The non-BPS black hole attractor equation'', JHEP 0603 (2006) 060, arXiv:hep-th/0602005; B. Sahoo and A. Sen, ``Higher derivative corrections to non-supersymmetric extremal black holes in $N=2$ supergravity'', JHEP 0609 (2006) 029, arXiv:hep-th/0603149; A. Dabholkar, A. Sen and S. P. Trivedi, ``Black hole microstates and attractor without supersymmetry'', JHEP 0701 (2007) 096, arXiv:hep-th/0611143; L. Andrianopoli, R. D'Auria, S. Ferrara and M. Trigiante, ``Extremal black holes in supergravity'', Lect. Notes Phys. 737 (2008) 661, arXiv:hep-th/0611345; G. L. Cardoso, B. de Wit and S. Mahapatra, ``Black hole entropy functions and attractor equations'', JHEP 0703 (2007) 085, arXiv:hep-th/0612225.

\bibitem{wald}
R. M. Wald, ``Black hole entropy is Noether charge'', Phys. Rev. D48 (1993) R3427, arXiv:gr-qc/9307038; T. Jacobson, G. Kang and R. C. Myers, ``On black hole entropy'', Phys. Rev. D49 (1994) 6587, arXiv:gr-qc/9312023; V. Iyer and R. M. Wald, ``Some properties of Noether charge and a proposal for dynamical black hole entropy'', Phys. Rev. D50 (1994) 846, arXiv:gr-qc/9403028; T. Jacobson, G. Kang and R. C. Myers, ``Black hole entropy in higher curvature gravity'', arXiv:gr-qc/9502009.

\bibitem{r2entropy}
G. L. Cardoso, B. de Wit and T. Mohaupt, ``Corrections to
macroscopic supersymmetric black-hole entropy'', Phys. Lett. B451
(1999) 309, arXiv:hep-th/9812082.

\bibitem{kastorwin}
D. Kastor and K. Z. Win, ``Non-extreme Calabi-Yau black holes'',
Phys. Lett. B411 (1997) 33, arXiv:hep-th/9705090.

\bibitem{nearextremal}
K. Behrndt, M. Cveti\v{c} and W. A. Sabra, ``Entropy of near-extreme
$N=2$ black holes'', Phys. Rev. D58 (1998) 084018,
arXiv:hep-th/9712221.

\bibitem{osv}
H. Ooguri, A. Strominger, C. Vafa, ``Black hole attractors and the
topological string'', Phys. Rev. D70 (2004) 106007,
arXiv:hep-th/0405146.

\bibitem{genstabeq}
W. A. Sabra, ``General static $N=2$ black holes'', Mod. Phys. Lett.
A12 (1997) 2585, arXiv:hep-th/9703101; W. A. Sabra, ``Black holes in
$N=2$ supergravity theories and harmonic functions'', Nucl. Phys.
B510 (1998) 247, arXiv:hep-th/9704147; K. Behrndt, D. L\"ust and W.
A. Sabra, ``Stationary solutions of $N=2$ supergravity'', Nucl.
Phys. B510 (1998) 264, arXiv:hep-th/9705169.

\bibitem{yeom}
B. de Wit, ``$N=2$ Electric-magnetic duality in a chiral
background'', Nucl. Phys. Proc. Suppl. 49 (1996) 191,
arXiv:hep-th/9602060.

\bibitem{smallbh}
A. Dabholkar, R. Kallosh and A. Maloney, ``A stringy cloak for a
classical singularity'', JHEP 0412 (2004) 059, arXiv:hep-th/0410076.

\bibitem{barakol}
B. Kol, ``A new action-derived form of the black hole metric'',
arXiv:gr-qc/0608001.

\end{thebibliography}
\end{document}